\documentclass[aps,prb,groupedaddress,twocolumn]{revtex4-2}

\usepackage{booktabs}
\usepackage{siunitx}
\usepackage{amssymb}

\usepackage{graphicx}
\usepackage{amsmath}
\usepackage{amssymb}
\usepackage{hyperref}

\newcommand{\avg}[1]{\left<#1\right>} 
\renewcommand{\vec}[1]{\ensuremath{\mathbf{#1}}} 

\begin{document}

\title{Low-spin ground state of the giant single-molecule magnets \{Mn$_{70}$\} and \{Mn$_{84}$\}}

\author{Roman Rausch}
\author{Christoph Karrasch}
\affiliation{Technische Universit\"at Braunschweig, Institut f\"ur Mathematische Physik, Mendelssohnstraße 3,
38106 Braunschweig, Germany}




\begin{abstract}
The single-molecule magnets \{Mn$_{70}$\} and \{Mn$_{84}$\} are characterized by a 14-site unit cell with $S=2$ spin sites arranged in a circular geometry. Experimentally, these systems exhibit a magnetic ground state with a notably low total spin $S_{\text{tot}}=5-7$. Up to now, this low-spin ground state has been up difficult to describe theoretically due to the complexity of the quantum Heisenberg model for such a large system. In this work, we fill this gap and demonstrate that the ground state of \{Mn$_{70}$\} and \{Mn$_{84}$\} is in fact governed by a small, finite $S_{\text{tot}}$ in quantitative agreement with the experiment. We employ accurate, large-scale SU(2)-symmetric density-matrix renormalization group calculations for a quantum Heisenberg model with previously published exchange parameters obtained by density-functional theory. We do not find a low-spin state for the same parameters and $S=1$ and thus propose that frustrated systems with $S\geq2$ are inherently prone to weak ferromagnetic interactions.  This could account for the prevalence of similar low-spin Mn-based single-molecule magnets. Finally, we compute the full magnetization curve and find wide plateaus at 10/14, 11/14, 12/14 and 13/14 of the saturation, which can be traced back to nearly-independent 3-site clusters with broken inter-cluster bonds.
\end{abstract}

\maketitle


\section{Introduction}

\begin{figure*}
\begin{center}
\includegraphics[width=0.6\textwidth]{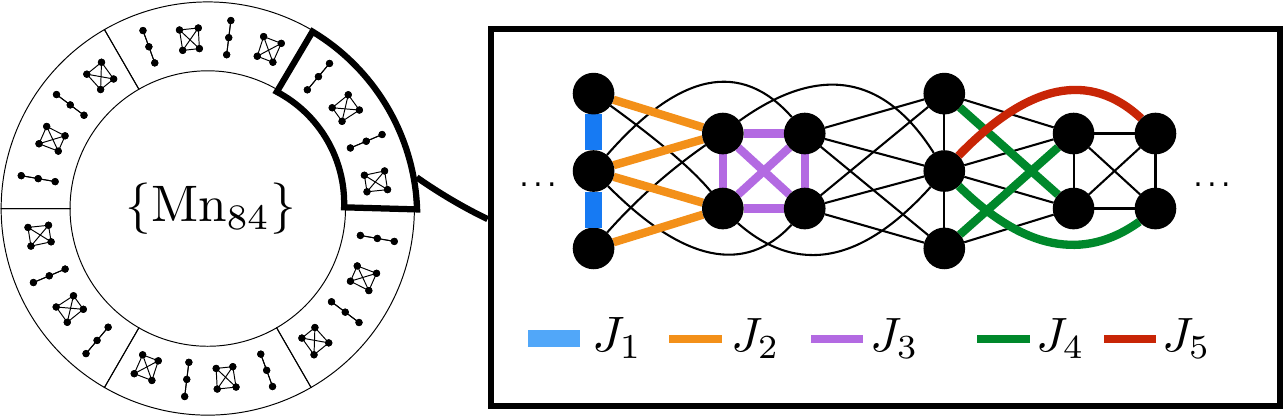}
\caption{
\label{fig:MnLattice}
\textit{Left:} Schematic picture of the spin sites in \{Mn$_{84}$\}. \textit{Right:} The 14-site unit cell of the Mn wheels, consisting out of the 3-site line piece and the 4-site tetrahedron. The first five DFT-calculated exchange interactions are (in units of meV): $J_1=13$, $J_2=3.2$, $J_3=1.1$, $J_4=0.4$, $J_5=-1.7$. Only $J_5$ is ferromagnetic. The couplings are periodically extended, but $J_1$-$J_3$ ($J_4$, $J_5$) are only highlighted in color within the first (second) 7-site segment.
}
\end{center}
\end{figure*}

Recent decades saw significant advances in creating large magnetic aggregates~\cite{Papatriantafyllopoulou2016}, which have evolved beyond zero-dimensional molecules and can now often be considered as finite but large condensed-matter systems. A particular highlight is the family of Mn wheel molecules, which have a unit cell of $L_{\text{cell}}=14$ magnetic centres arranged on a ring (see Fig.~\ref{fig:MnLattice}). The spin sites are Mn$^{3+}$ ions with $S=2$.
The successfully synthesized variants are \{Mn$_{70}$\}~\cite{Vinslava2016} and \{Mn$_{84}$\}~\cite{Tasiopoulos2004,Papatriantafyllopoulou2016} with $N_{\text{cells}}=5, 6$ unit cells and $L=L_{\text{cell}}\cdot N_{\text{cells}}$ total sites. The variants \{Mn$_{56}$\} and \{Mn$_{98}$\} ($N_{\text{cells}}=4,7$) might also be fabricable~\cite{Vinslava2016}. The \{Mn$_{84}$\} wheel has set the record as the largest single-molecule magnet (SMM)~\cite{Papatriantafyllopoulou2016}. Despite the SMM classification, it actually features a relatively small ground-state spin of $S_{\text{tot}} = 6$, i.e., $S_{\text{tot}}/L\approx0.07$,  $S_{\text{tot}}/S_{\text{max}}\approx0.04$ with $S_{\text{max}}=SL=2L$. Similarly, $S_{\text{tot}} = 5-8$ ($S_{\text{tot}}/L\approx0.07\text{--}0.11$, $S_{\text{tot}}/S_{\text{max}}\approx0.04\text{--}0.06$) is found for \{Mn$_{70}$\}~\cite{Vinslava2016} depending on the experimental extraction procedure and the precise chemical environment. In fact, such a low-spin state is quite typical among Mn-based SMMs. Reported compounds are listed in Tab.~\ref{tab:MnComplexes}.

\begin{table*}
\centering
\begin{tabular}{l l l c c l}
\toprule
$L$ & geometry features & spin values/max. spin & $S_{\text{tot}}$ & $S_{\text{tot}}/S_{\text{max}}$ & Ref.\\
\midrule
30 & C$_2$, near-octahedral & $3\cdot5/2+26\cdot2+3/2=61$ & 5-7 & 0.08-0.11 & \cite{Soler2001,Soler2004}\\
32 & mostly octahedral & $18\cdot5/2+10\cdot2+4\cdot3/2=71$ & 5 & 0.07 & \cite{Langley2011}\\
32 & truncated cube & $24\cdot5/2+8\cdot3/2=72$ & 9--10 & 0.13--0.14 & \cite{Scott2005}\\
32 & ``double-decker wheel'' & $18\cdot5/2+14\cdot2=73$ & 11--12 & 0.15--0.16 & \cite{Manoli2011}\\
40 & ``loop of loops'' & $8\cdot 5/2+32\cdot 2=84$ & 4 & 0.05 & \cite{Moushi2007}\\
70 & giant wheel & $70\cdot2=140$ & 5--8 & 0.04--0.06 & \cite{Vinslava2016}\\
84 & giant wheel & $84\cdot2=164$ & 6 & 0.04 & \cite{Tasiopoulos2004}\\
\bottomrule
\end{tabular}
\caption{\label{tab:MnComplexes}
Selection of low-spin molecular complexes that have Mn atoms as magnetic centres.
}
\end{table*}

In the absence of anisotropies, localized spins are described by the isotropic Heisenberg model
\begin{equation}
H = \sum_{ij} J_{ij} \vec{S}_i \cdot \vec{S}_j - g\mu_B B \sum_i S^z_i,
\label{eq:H}
\end{equation}
where $\vec{S}_i=(S^x_i,S^y_i,S^z_i)$ is the vector of spin operators at site $i$, $g$ is the gyromagnetic ratio, $\mu_B$ is the Bohr magneton, and $B$ is the external field. For a system of $L$ spins, the interactions are defined by the $L\times L$ matrix $J_{ij}$, which is best conveyed graphically (see Fig.~\ref{fig:MnLattice}). For $B=0$, the system has SU(2) invariance and the total spin
\begin{equation}
\sum_{ij} \left\langle\vec{S}_i\cdot\vec{S}_j\right\rangle = \avg{\vec{S}_{\text{tot}}^2} = S_{\text{tot}}\left(S_{\text{tot}}+1\right)
\end{equation}
is a conserved quantity. The total magnetization is given by
\begin{equation}
M_{\text{tot}} = \sum_i \langle S^z_i\rangle
\end{equation}
and is a conserved quantity for $B\geq0$.

The theoretical description of large molecules where the number of sites approaches $L\sim100$ poses a challenge to theory. Twenty years ago, the benchmark of theory-experiment comparison was set by Mn$_{12}$Ac (four sites with $S=3/2$ and eight sites with $S=2$) whose Hilbert space size is $\sim8.6\cdot 10^6$ in the  subspace $M_{\text{tot}}=0$~\cite{Regnault2002,Park2004,Chaboussant2004}. This problem is amenable to Lanczos diagonalization, so that an extensive comparison with the experiment could be carried out, and the family of these systems was even termed the ``drosophila of single-molecule magnetism''~\cite{Bagai2009}. Lanczos diagonalization runs into problems in the case of \{Mn$_{84}$\} whose Hilbert space size is $\sim 1.6\cdot10^{57}$ for $M_{\text{tot}}=0$, and  no theoretical progress could be made.

Renewed interest in the \{Mn$_{84}$\} wheel from theory came sixteen years after its original discovery~\cite{Schurkus2020,Chen2022}. In a pioneering work, density-functional theory (DFT) was used to compute the three dominant exchange integrals $J_{1-3}$ and the system was subsequently studied in a simplified fashion~\cite{Schurkus2020}: The ground state of the \{Mn$_7$\} subunit was discussed and the full system was treated by coupling coarse-grained subunits. The polarized ground state could not be explained because all the three exchange integrals turned out to be antiferromagnetic, but it was speculated that such a state should arise for longer-ranged ferromagnetic interactions. This approach was refined in a pivotal follow-up work~\cite{Chen2022}, where the first seven exchange integrals $J_{1-7}$ were computed, and $J_5$ turned out to be ferromagnetic (see Fig.~\ref{fig:MnLattice}). The specific heat and the magnetization curve were computed in terms of the classical Ising and Potts models by using a geometry-adapted tensor-network representation of the partition function. Up to this point, an accurate treatment of the full quantum Heisenberg model and a clear verification of the low-spin ground state of \{Mn$_{84}$\} is still missing.

On a different front, theoretical methods have been pushed in the last years to solve the full problem of large clusters of highly interconnected isotropic quantum spins. E.g., by using large-scale density-matrix renormalization group (DMRG) calculations, the long-standing problem of understanding the ground state on the pyrochlore lattice was solved for finite clusters with $L=64-128$ spins ($S=1/2,1$)~\cite{Hagymasi2021,Hagymasi2022}. The caveat is that the restriction to a finite system is still an uncontrolled approximation for lattices (especially in higher dimensions), but no such problem exists for molecules, where a conclusive solution was obtained for up to $L=60$ spins ($S=1/2$)~\cite{Rausch2021,Rausch2022}. A novel competing approach comes from a method based on neural networks~\cite{Szabo2024}.

In this paper, we employ large-scale DMRG numerics to significantly extend the previous classical-spin results for \{Mn$_{84}$\} with $S=2$. We show that by exploiting the full SU(2) symmetry of the problem, the full quantum-mechanical Heisenberg model remains tracatable and energy gaps of $\sim10^{-4}J_1$ can be resolved through effective bond dimensions of the order of $\chi_\text{eff}\sim50000$. We use the published set of exchange integrals~\cite{Chen2022}, but neglect the small long-range values $J_6$ and $J_7$ which complicate the numerics. We demonstrate that the remaining five exchange integrals~\cite{Chen2022} quantitatively reproduce the low-spin ground state $S_{\text{tot}}/N_{\text{cells}}\approx 1$ without any additional fitting, thereby closing the gap between theory and experiment.

\section{\label{sec:technical}Technical details}
 
The DMRG algorithm approximates the ground-state wavefunction as a matrix-product state, capturing entanglement and correlations through iterative optimization and state truncation. The key control parameter is the bond dimension $\chi$ that corresponds to the number of variational parameters in the ansatz state. As a rigorous measure of accuracy, we employ the variance per site given by
\begin{equation}
\text{var}/L = \bigg|\avg{\frac{H^2}{J_1^2}}-\avg{\frac{H}{J_1}}^2\bigg|/L.
\label{eq:var}
\end{equation}
We extrapolate the energy per site by a linear ansatz $E/L = (E/L)_{\text{exact}} + \alpha\cdot\text{var}/L$~\cite{Rausch2023}.
We use the two-site DMRG algorithm to grow the bond dimension for the first iterations before switching to the cheaper one-site algorithm with perturbations~\cite{Hubig2015}. The system naturally features periodic boundary conditions, which we implement via long-range terms on the level of the matrix-product operator representation of the Hamiltonian, and the bandwidth of the interaction graph $J_{ij}$ is minimized using the reverse Cuthill-McKee algorithm~\cite{Ummethum2013,Rausch2021}.

The SU(2) symmetry of the model~\eqref{eq:H} with $B=0$ allows us to increase efficiency by working in the reduced spin basis. A SU(2)-invariant bond dimension $\chi_{\text{SU(2)}}$ corresponds to a much larger effective $\chi_{\text{eff}}$. For the $S=2$ system investigated here, we find $\chi_{\text{eff}} \sim 7\chi_{\text{SU(2)}}$, i.e., almost an order-of-magnitude gain. Moreover, the exploitation of SU(2) symmetry allows us to directly target states with a given quantum number $S_{\text{tot}}$ and compute the energy curve $E(S_{\text{tot}})$.

As a guiding principle, we will also study a system with an infinite number of unit cells $N_\text{cells}\to\infty$. To this end, we employ the variational uniform matrix-product state (VUMPS) formalism \cite{Zauner-Stauber2018} with a numerical unit cell equal to the Mn$_{14}$ subunit. As a state error analogous to Eq.~\eqref{eq:var}, we use the two-site variance, see Ref.~\cite{Zauner-Stauber2018} for details. Within the VUMPS algorithm, we can exploit the SU(2) symmetry but can only access the sector $S_{\text{tot}}=0$ which does not necessarily correspond to the ground in frustrated systems with mixed antiferromagnetic and ferromagnetic interactions where the Lieb-Mattis theorem~\cite{Lieb_Mattis_1962} does not hold. One can also exploit only the U(1) symmetry, but this approach restricts polarizations within the unit cell to integer values. If no symmetry is exploited, the algorithm tends to converge to the lowly entangled sector with $M_{\text{tot}}=S_{\text{tot}}$, and we can directly compute the total spin per site~\cite{Rausch2023}, which might take irrational values. In a nutshell, exploiting different symmetries allows us to perform a cross-diagnostic analysis of the systems.

\begin{figure}[b]
\begin{center}
\includegraphics[width=\columnwidth]{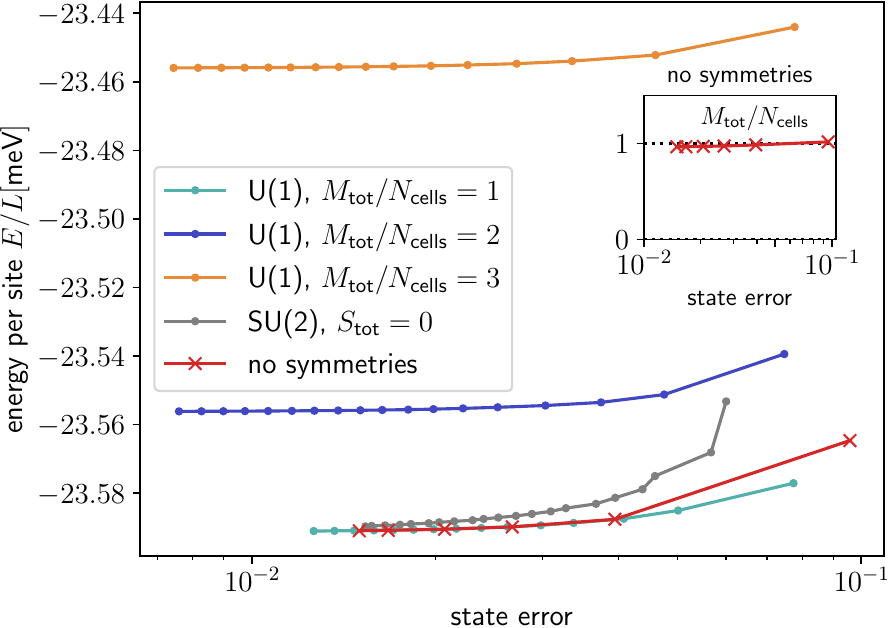}
\caption{
\label{fig:Mn84_VUMPS}
VUMPS data for the energy per site for an infinite number of coupled Mn$_{14}$ units at $B=0$ as a function of the state error that decreases with increasing bond dimension (for the definition see~Ref.~\cite{Zauner-Stauber2018}). Various symmetries are exploited in the algorithm, and the final bond dimensions reach $\chi = 500$ (no symmetries), $\chi \sim 2000$ [U(1) only], and $\chi_{\text{SU(2)}} \sim 3000$ [full SU(2)]. The inset shows the spin per number of unit cells for the case without symmetries; a linear fit gives the extrapolated value of $M_{\text{tot}}/N_{\text{cells}}\approx 0.96$. }
\end{center}
\end{figure}

\section{Infinite number of unit cells}

As a guiding principle for \{Mn$_{84}$\}, we first look at an infinite number of unit cells. Figure~\ref{fig:Mn84_VUMPS} shows the energies as a function of the state error which decreases with increasing bond dimension, see Ref.~\cite{Zauner-Stauber2018}.
The VUMPS calculation that does not exploit any symmetries yields the lowest energy and a polarization of $M_{\text{tot}}/N_{\text{cells}}\approx 0.96$. This energy is very close to the one obtained from a U(1) symmetric calculation in the sector $M_{\text{tot}}/N_{\text{cells}}=1$. The spin singlet $S_\text{tot}=0$ accessed by exploiting SU(2) symmetry is also close in energy, but remains slightly above. Higher integer polarizations $M_{\text{tot}}/N_{\text{cells}}>1$ within the 14-site unit cell clearly correspond to excited states. Note that a U(1)-symmetric calculation with $M_\text{tot}=0$ yields no new information: For a ground state with $S_\text{tot}=0$, the results coincide with those of a SU(2)-symmetric calculation, and for one with $S_\text{tot}>0$, a degenerate state with $M_\text{tot}>0$ will have lower entanglement.

We conclude that the ground state of the system features an irrational polarization close to, but not exactly equal to $M_{\text{tot}}/N_{\text{cells}}=1$, with a small energy barrier towards the singlet. For finite systems up to $L=84$ where the total spin can only change in integer steps, we expect a ground state with $S_{\text{tot}}/N_{\text{cells}}=1$, i.e., the integer spin closest to $S_{\text{tot}}/N_{\text{cells}}\approx 0.96$.

\section{Finite rings}
 
\begin{figure*}
\begin{center}
\includegraphics[width=0.75\textwidth]{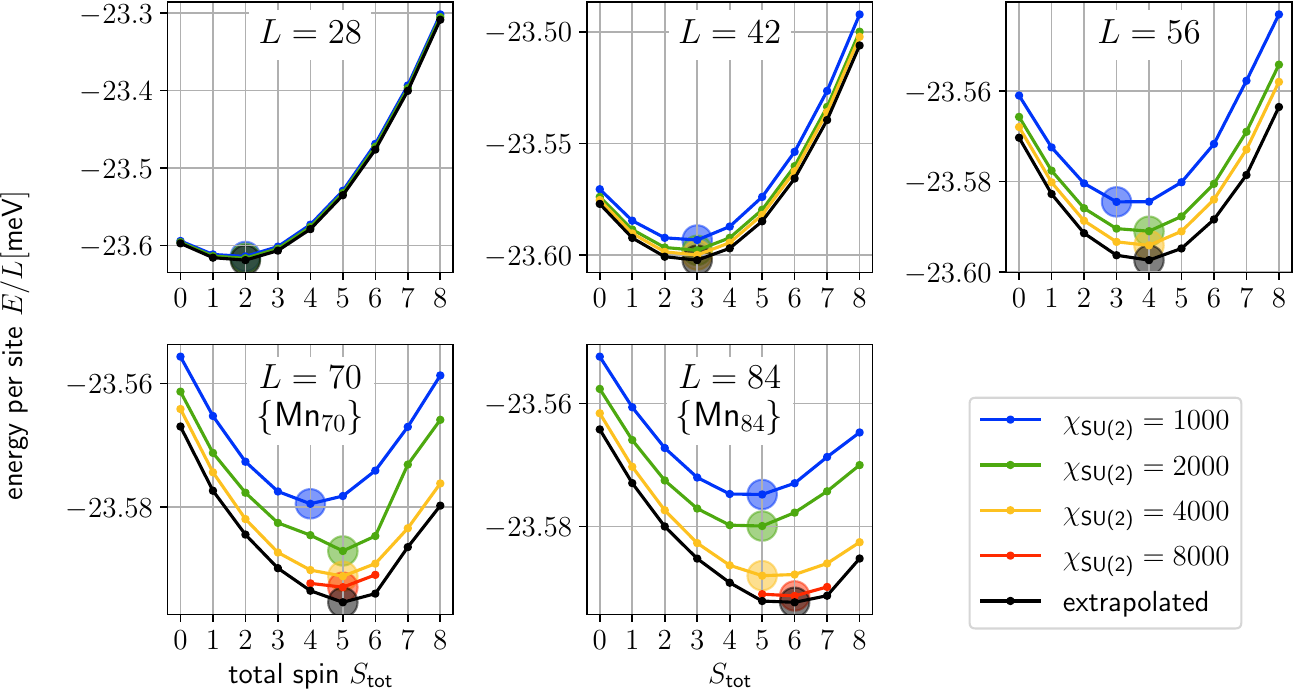}
\caption{
\label{fig:Mn84_towers}
Lowest energies per site (spin) in sectors with a fixed total spin $S_{\text{tot}}$ at $B=0$ for rings with $N_{\text{cells}}=2-6$ unit cells (corresponding to $L=28,42,56,70,84$). The data was obtained using a SU(2) symmetric DMRG algorithm for various bond dimensions $\chi_{\text{SU(2)}}$. The black dots are linearly extrapolated in the variance per site, see Sec.~\ref{sec:technical}. A circle highlights the energy minimum for each curve. The true ground state (overall minimum) features $S_{\text{tot}}/N_{\text{cells}}=1$ for all cases.
}
\end{center}
\end{figure*}

We now turn to finite rings of $N_{\text{cells}}=2-6$ unit cells with periodic boundary conditions; recall that $N_{\text{cells}}=6$ corresponds to \{Mn$_{84}$\}. We use the SU(2)-symmetric DMRG algorithm to compute the lowest energy in each spin sector in the range $S_{\text{tot}}=0-8$. This is shown in Fig.~\ref{fig:Mn84_towers}, which is the central result of our work.

For $N_{\text{cells}}=2,3,4$, a bond dimension of $\chi_{\text{SU(2)}} \leq 4000$ is sufficient to obtain converged data. The variance per site is of the order of $10^{-4}\text{meV}^2$ for $\chi_{\text{SU(2)}} = 4000$.
For $N_{\text{cells}}=5,6$, which correspond to the physical cases of \{Mn$_{70}$\} and \{Mn$_{84}$\}, the gaps become smaller and we have to go up to $\chi_{\text{SU(2)}} = 8000$ to obtain the same accuracy. In all cases, the true ground state features $S_{\text{tot}}/N_{\text{cells}}=1$.
However, $S_{\text{tot}}=5$ and $S_{\text{tot}}=6$ are extremely close in energy for \{Mn$_{84}$\}, and $S_{\text{tot}}=6$ only wins by a very small margin for $\chi_{\text{SU(2)}} = 8000$.
A conservative estimate is that $S_{\text{tot}}=5,6$ are near-degenerate for \{Mn$_{84}$\} with an energy per site of $E/(LJ_1) = -1.8148\pm0.0002$.

Overall, our large-scale DMRG calculations paired with exchange integrals from DFT decisively confirm the experimentally-measured total-spin state of \{Mn$_{84}$\}. We emphasize that in order to tackle this problem, it was imperative to employ a state-of-the art, SU(2) symmetric algorithm and bond dimensions of up to $\chi_{\text{SU(2)}} = 8000$ which corresponds to a much larger $\chi_\text{eff}\sim50000$. There may still be some influence from factors we have neglected, such as the inclusion of longer-ranged couplings or distortions in the interatomic distances. We hypothesize that these additional influences are small enough to preserve the general shape of the $S_{\text{tot}}/N_{\text{cells}}\approx 1$ minimum.

\section{Dependence on the spin quantum number}
 
\begin{figure}[t]
\begin{center}
\includegraphics[width=\columnwidth]{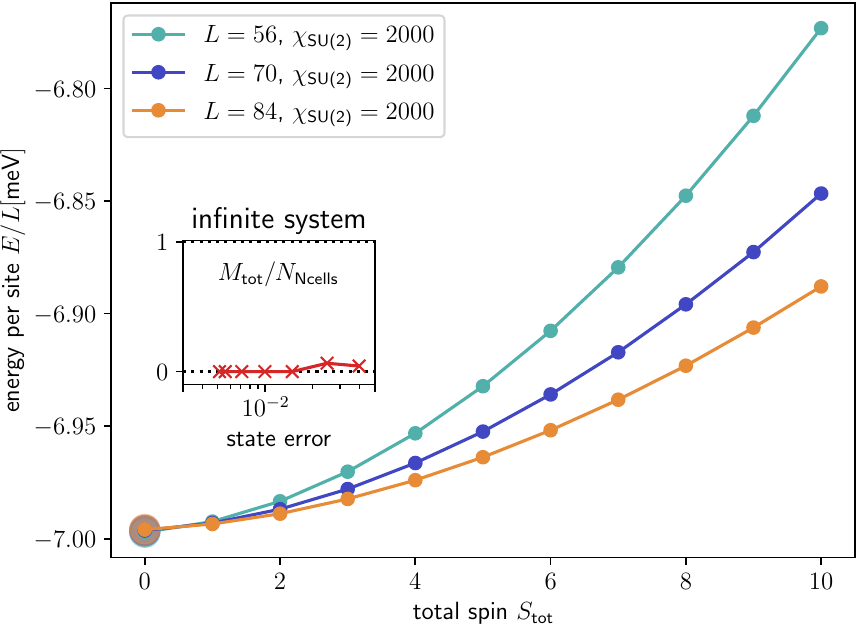}
\caption{
\label{fig:Mn84_tower_S=1}
Ground-state energies per site in different sectors of the total spin for a hypothetical system with $S=1$ on the same geometry and with the same exchange parameters as in Fig.~\ref{fig:MnLattice} for $L=56,70,84$ and $B=0$. All data are for $\chi_{\text{SU(2)}}=2000$. The true ground state always corresponds to a spin singlet $S_\text{tot}=0$. This is supported by the polarization in the infinite system (inset; VUMPS calculation, no symmetries).
}
\end{center}
\end{figure}

We briefly test whether a hypothetical $S=1$ system on the same geometry and with the same exchange parameters would also feature a ground state with a finite but small value of the total spin. Figure~\ref{fig:Mn84_tower_S=1} shows the lowest energy in different sectors of $S_\text{tot}$ for $L=56,70,84$. A bond dimension of $\chi_{\text{SU(2)}}=2000$ is sufficient to obtain a variance per site of the order of $10^{-5}\text{meV}^2$. We find that the overall minimum (i.e., the true ground state) is always at $S_{\text{tot}}=0$. The infinite-system calculation similarly shows $M_{\text{tot}}\approx0$ (inset to Fig.~\ref{fig:Mn84_tower_S=1}).
 
We conclude that for an $S=1$ system, the singlet ground state is robust against weak ferromagnetic perturbations, whereas it becomes susceptible to them in the case of $S=2$ and leads to a polarized ground state. Intuitively, it is clear that due to a large number of internal states, a high spin can ``afford'' a coexistence of a partial polarization and quantum fluctuations to minimize the energy, and in this sense becomes more classical. The non-trivial result is that this transition happens between $S=1$ and $S=2$.
This observation may explain the multitude of low-spin Mn-based SMMs listed in Tab.~\ref{tab:MnComplexes}, where the Mn sites have a local spin value that varies in the range $S=2\pm 0.5$ depending on the oxidation state (Mn$^{2+}$ to Mn$^{4+}$).

\section{Magnetization curve}
 
In this section we compute the magnetization curve for finite magnetic fields. We can determine the ground state of $H$ by first computing the lowest-energy eigenstates $|S_\text{tot}\rangle$ at $B=0$ for all values of $S_\text{tot}=0...2L$ using a SU(2)-symmetric algorithm. These are also eigenstates at finite $B$, and the overall ground state can be obtained by minimizing
\begin{equation}
\langle S_\text{tot} |H(B=0)|S_\text{tot}\rangle - g\mu_BB S_{\text{tot}}
\end{equation}
w.r.t.~$S_\text{tot}$; note that the second term in Eq.~(\ref{eq:H}) becomes minimal for $M_\text{tot}=S_\text{tot}$. Since this is a large volume of computations, we limit ourselves to $\chi_{\text{SU(2)}}=1000$. However, the full spectrum is on an energy scale of $>4500\text{meV}$, so that sub-meV energy inaccuracies carry no weight in this case. Moreover, the reduced Hilbert space at high spins makes the calculations easier and we find that all the energies with $S_{\text{tot}} \gtrsim 114$ for $L=84$ are in fact numerically exact for this bond dimension.
 
\begin{figure}[b]
\begin{center}
\includegraphics[width=\columnwidth]{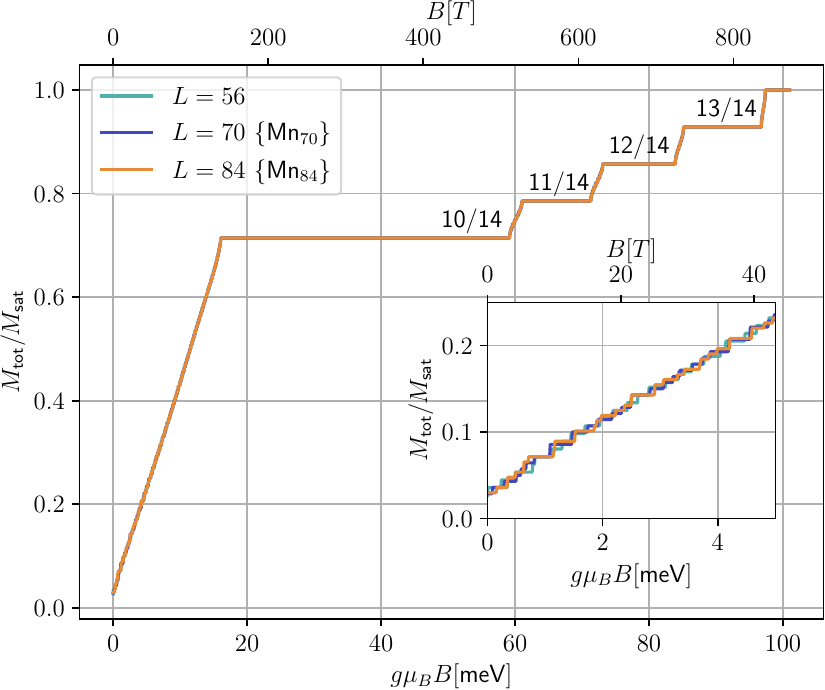}
\caption{
\label{fig:Mn84_mag}
Magnetization $M_\text{tot}$ as a function of the magnetic field $B$ for $L=56,70,84$ at a fixed bond dimension $\chi_{\text{SU(2)}}=1000$. The curves are normalized to the saturation value $M_\text{sat}=2L$ and are practically on top of each other on this scale. The inset shows a zoom-in on the low-field onset.
}
\end{center}
\end{figure}

Figure~\ref{fig:Mn84_mag} shows the magnetization $M_\text{tot}$ as a function of the magnetic field for $L=56,70,84$. Starting from the zero-field limit $M_\text{tot}(B\to0)=N_\text{cells}$, we observe a linear increase (with small finite-size steps) until $g\mu_B B\sim 16\text{meV}$, i.e., until the field becomes larger than the dominant exchange $J_1=13\text{meV}$. This is followed by a wide plateau at 10/14 of the saturation, and again by three more narrow plateaus at 11/14, 12/14, and 13/14. We note that due to the large value of $J_1$ (which corresponds to $T\approx151\text{K}/k_B$), even the first plateau will be quite difficult to observe, requiring at least $B\sim 160\text{T}$.

\begin{figure*}[t]
\begin{center}
\includegraphics[width=0.65\textwidth]{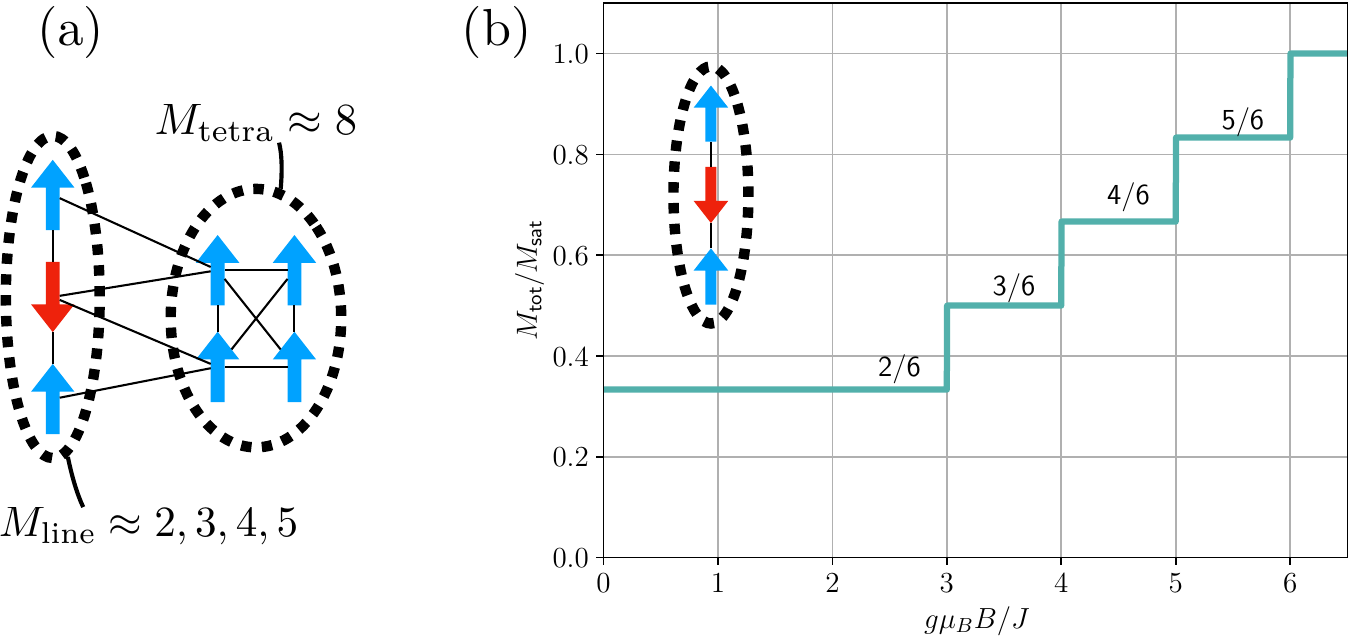}
\caption{
\label{fig:MnPlateaus}
a) Schematic representation of the local polarization at the magnetization plateaus. The 10/14, 11/14, 12/14, 13/14 plateaus correspond to magnetization values of the line piece of $M_{\text{line}}=2,3,4,5$, while the tetrahedron is nearly fully polarized.
b) Magnetization curve of a 3-spin line piece with $S=2$.
}
\end{center}
\end{figure*}

In order to better understand the origin of the plateaus, we compute the site-resolved magnetizations $\avg{S^z_i}$ and find that the tetrahedra at these plateaus are almost completely polarized with $M_{\text{tetra}}\approx 8$. This is understandable because the coupling $J_3=1.1\text{meV}$ is weak.
Since the line pieces are coupled via the tetrahedra (see Fig.~\ref{fig:MnLattice}), they disconnect and the physics should be governed by an ensemble of independent 3-site clusters described by the Hamiltonian
\begin{equation}
H_{\text{eff}} = J \left( \vec{S}_1\cdot\vec{S}_2 + \vec{S}_2\cdot\vec{S}_3 \right) - g\mu_BB \sum_{i=1}^3 S^z_i.
\end{equation}
To support this hypothesis, we compute the magnetization curve of $H_{\text{eff}}$ for $S=2$ [see Fig.~\ref{fig:MnPlateaus} b)] and find a 2/6 plateau of width $g\mu_B\Delta B=3J$ as well as three plateaus at 3/6, 4/6, 5/6 of width $g\mu_B\Delta B=J$. The central spin aligns itself antiparallel to the others to maximize the exchange energy and is flipped in a stepwise fashion.
This quantitatively explains the number and widths of the plateaus for the full case in Fig.~\ref{fig:Mn84_mag} if one sets $J=J_1=13\text{meV}$ and rescales the fraction of the saturation by including the tetrahedra $M/M_{\text{sat}}=(n+8)/(6+8)$, $n=2,3,4,5$.

Magnetization plateaus are typically related to localized magnons, where spinflips localize due to a canceling of hopping terms on a frustrated geometry~\cite{Schnack2001,Schulenburg2002,Richter2004,Richter2005}. In our case, the reason for localization is the inhomogeneous form of the interactions; the external field divides the system into independent islands by breaking the weak bonds that connect them. A similar effect was observed in Atacamite \cite{heinze2024atacamitecu2cloh3highmagnetic}.

\section{Conclusion}

We have studied the quantum Heisenberg model on the geometry of the Mn giant wheels using a set of DFT-computed exchange parameters~\cite{Chen2022} combined with large-scale SU(2)-symmetric DMRG calculations where the effective bond dimension reaches values of $\chi_\text{eff}\sim50000$. The presence of a unit cell allows us to scale the system size from finite rings with $N_{\text{cells}}=2-6$ and also to compare with the infinite chain as a guiding case. The ground state clearly features a total spin $S_{\text{tot}}/N_{\text{cells}}\approx 1$, which quantitatively reproduces the experimental values ($S_{\text{tot}}=6$ for \{Mn$_{84}$\}, $N_{\text{cells}}=6$) without any additional free parameters. This shows that theory has caught up with experimental progress in manufacturing high-nucleation SMMs and is able to successfully describe large finite magnetic systems with $L\sim100$ spin sites.

By reducing the spin quantum number to $S=1$, we find that the ground state robustly remains at $S_{\text{tot}}=0$. We hence argue that systems with $S\gtrsim2$ become susceptible to the presence of small ferromagnetic exchange interactions and react by a partial ground-state polarization. This explains the multitude of low-spin SMMs that are based on Mn ions and marks a crossover point from quantum to classical spins; the low-spin state emerges from a delicate interplay of three key factors, all of which are essential: i) the presence of a small ferromagnetic coupling,  ii) frustration, and iii) a spin $S>1$.

Finally, we find magnetization plateaus at 10/14, 11/14, 12/14 and 13/14 of the saturation; they are a consequence of the inhomogeous interactions in the system and can be explained by independent 3-site clusters whose weak inter-cluster couplings are broken by the external field.

\subsection*{Acknowledgements}
C.K. and R.R. acknowledge support by the Deutsche Forschungsgemeinschaft (DFG, German Research Foundation) under Germany's Excellence Strategy EXC-2123 QuantumFrontiers 390837967.


\bibliography{Mn84.bib}

\begin{thebibliography}{30}%
\makeatletter
\providecommand \@ifxundefined [1]{%
 \@ifx{#1\undefined}
}%
\providecommand \@ifnum [1]{%
 \ifnum #1\expandafter \@firstoftwo
 \else \expandafter \@secondoftwo
 \fi
}%
\providecommand \@ifx [1]{%
 \ifx #1\expandafter \@firstoftwo
 \else \expandafter \@secondoftwo
 \fi
}%
\providecommand \natexlab [1]{#1}%
\providecommand \enquote  [1]{``#1''}%
\providecommand \bibnamefont  [1]{#1}%
\providecommand \bibfnamefont [1]{#1}%
\providecommand \citenamefont [1]{#1}%
\providecommand \href@noop [0]{\@secondoftwo}%
\providecommand \href [0]{\begingroup \@sanitize@url \@href}%
\providecommand \@href[1]{\@@startlink{#1}\@@href}%
\providecommand \@@href[1]{\endgroup#1\@@endlink}%
\providecommand \@sanitize@url [0]{\catcode `\\12\catcode `\$12\catcode
  `\&12\catcode `\#12\catcode `\^12\catcode `\_12\catcode `\%12\relax}%
\providecommand \@@startlink[1]{}%
\providecommand \@@endlink[0]{}%
\providecommand \url  [0]{\begingroup\@sanitize@url \@url }%
\providecommand \@url [1]{\endgroup\@href {#1}{\urlprefix }}%
\providecommand \urlprefix  [0]{URL }%
\providecommand \Eprint [0]{\href }%
\providecommand \doibase [0]{https://doi.org/}%
\providecommand \selectlanguage [0]{\@gobble}%
\providecommand \bibinfo  [0]{\@secondoftwo}%
\providecommand \bibfield  [0]{\@secondoftwo}%
\providecommand \translation [1]{[#1]}%
\providecommand \BibitemOpen [0]{}%
\providecommand \bibitemStop [0]{}%
\providecommand \bibitemNoStop [0]{.\EOS\space}%
\providecommand \EOS [0]{\spacefactor3000\relax}%
\providecommand \BibitemShut  [1]{\csname bibitem#1\endcsname}%
\let\auto@bib@innerbib\@empty
\bibitem [{\citenamefont {Papatriantafyllopoulou}\ \emph
  {et~al.}(2016)\citenamefont {Papatriantafyllopoulou}, \citenamefont {Moushi},
  \citenamefont {Christou},\ and\ \citenamefont
  {Tasiopoulos}}]{Papatriantafyllopoulou2016}%
  \BibitemOpen
  \bibfield  {author} {\bibinfo {author} {\bibfnamefont {C.}~\bibnamefont
  {Papatriantafyllopoulou}}, \bibinfo {author} {\bibfnamefont {E.~E.}\
  \bibnamefont {Moushi}}, \bibinfo {author} {\bibfnamefont {G.}~\bibnamefont
  {Christou}},\ and\ \bibinfo {author} {\bibfnamefont {A.~J.}\ \bibnamefont
  {Tasiopoulos}},\ }\bibfield  {title} {\bibinfo {title} {Filling the gap
  between the quantum and classical worlds of nanoscale magnetism: giant
  molecular aggregates based on paramagnetic 3d metal ions},\ }\href
  {https://doi.org/10.1039/C5CS00590F} {\bibfield  {journal} {\bibinfo
  {journal} {Chem. Soc. Rev.}\ }\textbf {\bibinfo {volume} {45}},\ \bibinfo
  {pages} {1597} (\bibinfo {year} {2016})}\BibitemShut {NoStop}%
\bibitem [{\citenamefont {Vinslava}\ \emph {et~al.}(2016)\citenamefont
  {Vinslava}, \citenamefont {Tasiopoulos}, \citenamefont {Wernsdorfer},
  \citenamefont {Abboud},\ and\ \citenamefont {Christou}}]{Vinslava2016}%
  \BibitemOpen
  \bibfield  {author} {\bibinfo {author} {\bibfnamefont {A.}~\bibnamefont
  {Vinslava}}, \bibinfo {author} {\bibfnamefont {A.~J.}\ \bibnamefont
  {Tasiopoulos}}, \bibinfo {author} {\bibfnamefont {W.}~\bibnamefont
  {Wernsdorfer}}, \bibinfo {author} {\bibfnamefont {K.~A.}\ \bibnamefont
  {Abboud}},\ and\ \bibinfo {author} {\bibfnamefont {G.}~\bibnamefont
  {Christou}},\ }\bibfield  {title} {\bibinfo {title} {Molecules at the
  quantum--classical nanoparticle interface: Giant {Mn}$_{70}$ single-molecule
  magnets of $\sim$4 nm diameter},\ }\href
  {https://doi.org/10.1021/acs.inorgchem.5b02790} {\bibfield  {journal}
  {\bibinfo  {journal} {Inorganic Chemistry}\ }\textbf {\bibinfo {volume}
  {55}},\ \bibinfo {pages} {3419} (\bibinfo {year} {2016})}\BibitemShut
  {NoStop}%
\bibitem [{\citenamefont {Tasiopoulos}\ \emph {et~al.}(2004)\citenamefont
  {Tasiopoulos}, \citenamefont {Vinslava}, \citenamefont {Wernsdorfer},
  \citenamefont {Abboud},\ and\ \citenamefont {Christou}}]{Tasiopoulos2004}%
  \BibitemOpen
  \bibfield  {author} {\bibinfo {author} {\bibfnamefont {A.~J.}\ \bibnamefont
  {Tasiopoulos}}, \bibinfo {author} {\bibfnamefont {A.}~\bibnamefont
  {Vinslava}}, \bibinfo {author} {\bibfnamefont {W.}~\bibnamefont
  {Wernsdorfer}}, \bibinfo {author} {\bibfnamefont {K.~A.}\ \bibnamefont
  {Abboud}},\ and\ \bibinfo {author} {\bibfnamefont {G.}~\bibnamefont
  {Christou}},\ }\bibfield  {title} {\bibinfo {title} {Giant single-molecule
  magnets: A {Mn}$_{84}$ torus and its supramolecular nanotubes},\ }\href
  {https://doi.org/https://doi.org/10.1002/anie.200353352} {\bibfield
  {journal} {\bibinfo  {journal} {Angewandte Chemie International Edition}\
  }\textbf {\bibinfo {volume} {43}},\ \bibinfo {pages} {2117} (\bibinfo {year}
  {2004})},\ \Eprint
  {https://arxiv.org/abs/https://onlinelibrary.wiley.com/doi/pdf/10.1002/anie.200353352}
  {https://onlinelibrary.wiley.com/doi/pdf/10.1002/anie.200353352} \BibitemShut
  {NoStop}%
\bibitem [{\citenamefont {Soler}\ \emph {et~al.}(2001)\citenamefont {Soler},
  \citenamefont {Rumberger}, \citenamefont {Folting}, \citenamefont
  {Hendrickson*},\ and\ \citenamefont {Christou*}}]{Soler2001}%
  \BibitemOpen
  \bibfield  {author} {\bibinfo {author} {\bibfnamefont {M.}~\bibnamefont
  {Soler}}, \bibinfo {author} {\bibfnamefont {E.}~\bibnamefont {Rumberger}},
  \bibinfo {author} {\bibfnamefont {K.}~\bibnamefont {Folting}}, \bibinfo
  {author} {\bibfnamefont {D.~N.}\ \bibnamefont {Hendrickson*}},\ and\ \bibinfo
  {author} {\bibfnamefont {G.}~\bibnamefont {Christou*}},\ }\bibfield  {title}
  {\bibinfo {title} {Synthesis, characterization and magnetic properties of
  [mn30o24(oh)8(o2cch2c(ch3)3)32(h2o)2(ch3no2)4]: the largest manganese
  carboxylate cluster},\ }\href
  {https://doi.org/https://doi.org/10.1016/S0277-5387(01)00620-9} {\bibfield
  {journal} {\bibinfo  {journal} {Polyhedron}\ }\textbf {\bibinfo {volume}
  {20}},\ \bibinfo {pages} {1365} (\bibinfo {year} {2001})}\BibitemShut
  {NoStop}%
\bibitem [{\citenamefont {Soler}\ \emph {et~al.}(2004)\citenamefont {Soler},
  \citenamefont {Wernsdorfer}, \citenamefont {Folting}, \citenamefont {Pink},\
  and\ \citenamefont {Christou}}]{Soler2004}%
  \BibitemOpen
  \bibfield  {author} {\bibinfo {author} {\bibfnamefont {M.}~\bibnamefont
  {Soler}}, \bibinfo {author} {\bibfnamefont {W.}~\bibnamefont {Wernsdorfer}},
  \bibinfo {author} {\bibfnamefont {K.}~\bibnamefont {Folting}}, \bibinfo
  {author} {\bibfnamefont {M.}~\bibnamefont {Pink}},\ and\ \bibinfo {author}
  {\bibfnamefont {G.}~\bibnamefont {Christou}},\ }\bibfield  {title} {\bibinfo
  {title} {Single-molecule magnets:{\thinspace} a large {Mn}$_{30}$ molecular
  nanomagnet exhibiting quantum tunneling of magnetization},\ }\href
  {https://doi.org/10.1021/ja0297638} {\bibfield  {journal} {\bibinfo
  {journal} {Journal of the American Chemical Society}\ }\textbf {\bibinfo
  {volume} {126}},\ \bibinfo {pages} {2156} (\bibinfo {year}
  {2004})}\BibitemShut {NoStop}%
\bibitem [{\citenamefont {Langley}\ \emph {et~al.}(2011)\citenamefont
  {Langley}, \citenamefont {Stott}, \citenamefont {Chilton}, \citenamefont
  {Moubaraki},\ and\ \citenamefont {Murray}}]{Langley2011}%
  \BibitemOpen
  \bibfield  {author} {\bibinfo {author} {\bibfnamefont {S.~K.}\ \bibnamefont
  {Langley}}, \bibinfo {author} {\bibfnamefont {R.~A.}\ \bibnamefont {Stott}},
  \bibinfo {author} {\bibfnamefont {N.~F.}\ \bibnamefont {Chilton}}, \bibinfo
  {author} {\bibfnamefont {B.}~\bibnamefont {Moubaraki}},\ and\ \bibinfo
  {author} {\bibfnamefont {K.~S.}\ \bibnamefont {Murray}},\ }\bibfield  {title}
  {\bibinfo {title} {A high nuclearity mixed valence {Mn}$_{32}$ complex},\
  }\href {https://doi.org/10.1039/C1CC00035G} {\bibfield  {journal} {\bibinfo
  {journal} {Chem. Commun.}\ }\textbf {\bibinfo {volume} {47}},\ \bibinfo
  {pages} {6281} (\bibinfo {year} {2011})}\BibitemShut {NoStop}%
\bibitem [{\citenamefont {Scott}\ \emph {et~al.}(2005)\citenamefont {Scott},
  \citenamefont {Parsons}, \citenamefont {Murugesu}, \citenamefont
  {Wernsdorfer}, \citenamefont {Christou},\ and\ \citenamefont
  {Brechin}}]{Scott2005}%
  \BibitemOpen
  \bibfield  {author} {\bibinfo {author} {\bibfnamefont {R.~T.~W.}\
  \bibnamefont {Scott}}, \bibinfo {author} {\bibfnamefont {S.}~\bibnamefont
  {Parsons}}, \bibinfo {author} {\bibfnamefont {M.}~\bibnamefont {Murugesu}},
  \bibinfo {author} {\bibfnamefont {W.}~\bibnamefont {Wernsdorfer}}, \bibinfo
  {author} {\bibfnamefont {G.}~\bibnamefont {Christou}},\ and\ \bibinfo
  {author} {\bibfnamefont {E.~K.}\ \bibnamefont {Brechin}},\ }\bibfield
  {title} {\bibinfo {title} {Linking centered manganese triangles into larger
  clusters: A {Mn}$_{32}$ truncated cube},\ }\href
  {https://doi.org/https://doi.org/10.1002/anie.200501881} {\bibfield
  {journal} {\bibinfo  {journal} {Angewandte Chemie International Edition}\
  }\textbf {\bibinfo {volume} {44}},\ \bibinfo {pages} {6540} (\bibinfo {year}
  {2005})},\ \Eprint
  {https://arxiv.org/abs/https://onlinelibrary.wiley.com/doi/pdf/10.1002/anie.200501881}
  {https://onlinelibrary.wiley.com/doi/pdf/10.1002/anie.200501881} \BibitemShut
  {NoStop}%
\bibitem [{\citenamefont {Manoli}\ \emph {et~al.}(2011)\citenamefont {Manoli},
  \citenamefont {Inglis}, \citenamefont {Manos}, \citenamefont {Nastopoulos},
  \citenamefont {Wernsdorfer}, \citenamefont {Brechin},\ and\ \citenamefont
  {Tasiopoulos}}]{Manoli2011}%
  \BibitemOpen
  \bibfield  {author} {\bibinfo {author} {\bibfnamefont {M.}~\bibnamefont
  {Manoli}}, \bibinfo {author} {\bibfnamefont {R.}~\bibnamefont {Inglis}},
  \bibinfo {author} {\bibfnamefont {M.~J.}\ \bibnamefont {Manos}}, \bibinfo
  {author} {\bibfnamefont {V.}~\bibnamefont {Nastopoulos}}, \bibinfo {author}
  {\bibfnamefont {W.}~\bibnamefont {Wernsdorfer}}, \bibinfo {author}
  {\bibfnamefont {E.~K.}\ \bibnamefont {Brechin}},\ and\ \bibinfo {author}
  {\bibfnamefont {A.~J.}\ \bibnamefont {Tasiopoulos}},\ }\bibfield  {title}
  {\bibinfo {title} {A [{Mn}$_{32}$] double-decker wheel},\ }\href
  {https://doi.org/https://doi.org/10.1002/anie.201100976} {\bibfield
  {journal} {\bibinfo  {journal} {Angewandte Chemie International Edition}\
  }\textbf {\bibinfo {volume} {50}},\ \bibinfo {pages} {4441} (\bibinfo {year}
  {2011})},\ \Eprint
  {https://arxiv.org/abs/https://onlinelibrary.wiley.com/doi/pdf/10.1002/anie.201100976}
  {https://onlinelibrary.wiley.com/doi/pdf/10.1002/anie.201100976} \BibitemShut
  {NoStop}%
\bibitem [{\citenamefont {Moushi}\ \emph {et~al.}(2007)\citenamefont {Moushi},
  \citenamefont {Lampropoulos}, \citenamefont {Wernsdorfer}, \citenamefont
  {Nastopoulos}, \citenamefont {Christou},\ and\ \citenamefont
  {Tasiopoulos}}]{Moushi2007}%
  \BibitemOpen
  \bibfield  {author} {\bibinfo {author} {\bibfnamefont {E.~E.}\ \bibnamefont
  {Moushi}}, \bibinfo {author} {\bibfnamefont {C.}~\bibnamefont
  {Lampropoulos}}, \bibinfo {author} {\bibfnamefont {W.}~\bibnamefont
  {Wernsdorfer}}, \bibinfo {author} {\bibfnamefont {V.}~\bibnamefont
  {Nastopoulos}}, \bibinfo {author} {\bibfnamefont {G.}~\bibnamefont
  {Christou}},\ and\ \bibinfo {author} {\bibfnamefont {A.~J.}\ \bibnamefont
  {Tasiopoulos}},\ }\bibfield  {title} {\bibinfo {title} {A large
  [{Mn}$_{10}${Na}]$_4$ loop of four linked {Mn}$_{10}$ loops},\ }\href
  {https://doi.org/10.1021/ic062454o} {\bibfield  {journal} {\bibinfo
  {journal} {Inorganic Chemistry}\ }\textbf {\bibinfo {volume} {46}},\ \bibinfo
  {pages} {3795} (\bibinfo {year} {2007})}\BibitemShut {NoStop}%
\bibitem [{\citenamefont {Regnault}\ \emph {et~al.}(2002)\citenamefont
  {Regnault}, \citenamefont {Jolic\oe{}ur}, \citenamefont {Sessoli},
  \citenamefont {Gatteschi},\ and\ \citenamefont {Verdaguer}}]{Regnault2002}%
  \BibitemOpen
  \bibfield  {author} {\bibinfo {author} {\bibfnamefont {N.}~\bibnamefont
  {Regnault}}, \bibinfo {author} {\bibfnamefont {T.}~\bibnamefont
  {Jolic\oe{}ur}}, \bibinfo {author} {\bibfnamefont {R.}~\bibnamefont
  {Sessoli}}, \bibinfo {author} {\bibfnamefont {D.}~\bibnamefont {Gatteschi}},\
  and\ \bibinfo {author} {\bibfnamefont {M.}~\bibnamefont {Verdaguer}},\
  }\bibfield  {title} {\bibinfo {title} {Exchange coupling in the magnetic
  molecular cluster {Mn}$_{12}${Ac}},\ }\href
  {https://doi.org/10.1103/PhysRevB.66.054409} {\bibfield  {journal} {\bibinfo
  {journal} {Phys. Rev. B}\ }\textbf {\bibinfo {volume} {66}},\ \bibinfo
  {pages} {054409} (\bibinfo {year} {2002})}\BibitemShut {NoStop}%
\bibitem [{\citenamefont {Park}\ \emph {et~al.}(2004)\citenamefont {Park},
  \citenamefont {Pederson},\ and\ \citenamefont {Hellberg}}]{Park2004}%
  \BibitemOpen
  \bibfield  {author} {\bibinfo {author} {\bibfnamefont {K.}~\bibnamefont
  {Park}}, \bibinfo {author} {\bibfnamefont {M.~R.}\ \bibnamefont {Pederson}},\
  and\ \bibinfo {author} {\bibfnamefont {C.~S.}\ \bibnamefont {Hellberg}},\
  }\bibfield  {title} {\bibinfo {title} {Properties of low-lying excited
  manifolds in {Mn}$_{12}$ acetate},\ }\href
  {https://doi.org/10.1103/PhysRevB.69.014416} {\bibfield  {journal} {\bibinfo
  {journal} {Phys. Rev. B}\ }\textbf {\bibinfo {volume} {69}},\ \bibinfo
  {pages} {014416} (\bibinfo {year} {2004})}\BibitemShut {NoStop}%
\bibitem [{\citenamefont {Chaboussant}\ \emph {et~al.}(2004)\citenamefont
  {Chaboussant}, \citenamefont {Sieber}, \citenamefont {Ochsenbein},
  \citenamefont {G\"udel}, \citenamefont {Murrie}, \citenamefont {Honecker},
  \citenamefont {Fukushima},\ and\ \citenamefont {Normand}}]{Chaboussant2004}%
  \BibitemOpen
  \bibfield  {author} {\bibinfo {author} {\bibfnamefont {G.}~\bibnamefont
  {Chaboussant}}, \bibinfo {author} {\bibfnamefont {A.}~\bibnamefont {Sieber}},
  \bibinfo {author} {\bibfnamefont {S.}~\bibnamefont {Ochsenbein}}, \bibinfo
  {author} {\bibfnamefont {H.-U.}\ \bibnamefont {G\"udel}}, \bibinfo {author}
  {\bibfnamefont {M.}~\bibnamefont {Murrie}}, \bibinfo {author} {\bibfnamefont
  {A.}~\bibnamefont {Honecker}}, \bibinfo {author} {\bibfnamefont
  {N.}~\bibnamefont {Fukushima}},\ and\ \bibinfo {author} {\bibfnamefont
  {B.}~\bibnamefont {Normand}},\ }\bibfield  {title} {\bibinfo {title}
  {Exchange interactions and high-energy spin states in {Mn}$_{12}$-acetate},\
  }\href {https://doi.org/10.1103/PhysRevB.70.104422} {\bibfield  {journal}
  {\bibinfo  {journal} {Phys. Rev. B}\ }\textbf {\bibinfo {volume} {70}},\
  \bibinfo {pages} {104422} (\bibinfo {year} {2004})}\BibitemShut {NoStop}%
\bibitem [{\citenamefont {Bagai}\ and\ \citenamefont
  {Christou}(2009)}]{Bagai2009}%
  \BibitemOpen
  \bibfield  {author} {\bibinfo {author} {\bibfnamefont {R.}~\bibnamefont
  {Bagai}}\ and\ \bibinfo {author} {\bibfnamefont {G.}~\bibnamefont
  {Christou}},\ }\bibfield  {title} {\bibinfo {title} {The drosophila of
  single-molecule magnetism:
  [{Mn}$_{12}${O}$_{12}$({O}$_2${CR})$_{16}$({H}$_{2}${O})$_4$]},\ }\href
  {https://doi.org/10.1039/B811963E} {\bibfield  {journal} {\bibinfo  {journal}
  {Chem. Soc. Rev.}\ }\textbf {\bibinfo {volume} {38}},\ \bibinfo {pages}
  {1011} (\bibinfo {year} {2009})}\BibitemShut {NoStop}%
\bibitem [{\citenamefont {Schurkus}\ \emph {et~al.}(2020)\citenamefont
  {Schurkus}, \citenamefont {Chen}, \citenamefont {O'Rourke}, \citenamefont
  {Cheng},\ and\ \citenamefont {Chan}}]{Schurkus2020}%
  \BibitemOpen
  \bibfield  {author} {\bibinfo {author} {\bibfnamefont {H.~F.}\ \bibnamefont
  {Schurkus}}, \bibinfo {author} {\bibfnamefont {D.}~\bibnamefont {Chen}},
  \bibinfo {author} {\bibfnamefont {M.~J.}\ \bibnamefont {O'Rourke}}, \bibinfo
  {author} {\bibfnamefont {H.-P.}\ \bibnamefont {Cheng}},\ and\ \bibinfo
  {author} {\bibfnamefont {G.~K.-L.}\ \bibnamefont {Chan}},\ }\bibfield
  {title} {\bibinfo {title} {Exploring the magnetic properties of the largest
  single-molecule magnets},\ }\href
  {https://doi.org/10.1021/acs.jpclett.0c00020} {\bibfield  {journal} {\bibinfo
   {journal} {The Journal of Physical Chemistry Letters}\ }\textbf {\bibinfo
  {volume} {11}},\ \bibinfo {pages} {3789} (\bibinfo {year}
  {2020})}\BibitemShut {NoStop}%
\bibitem [{\citenamefont {Chen}\ \emph {et~al.}(2022)\citenamefont {Chen},
  \citenamefont {Helms}, \citenamefont {Hale}, \citenamefont {Lee},
  \citenamefont {Li}, \citenamefont {Gray}, \citenamefont {Christou},
  \citenamefont {Zapf}, \citenamefont {Chan},\ and\ \citenamefont
  {Cheng}}]{Chen2022}%
  \BibitemOpen
  \bibfield  {author} {\bibinfo {author} {\bibfnamefont {D.-T.}\ \bibnamefont
  {Chen}}, \bibinfo {author} {\bibfnamefont {P.}~\bibnamefont {Helms}},
  \bibinfo {author} {\bibfnamefont {A.~R.}\ \bibnamefont {Hale}}, \bibinfo
  {author} {\bibfnamefont {M.}~\bibnamefont {Lee}}, \bibinfo {author}
  {\bibfnamefont {C.}~\bibnamefont {Li}}, \bibinfo {author} {\bibfnamefont
  {J.}~\bibnamefont {Gray}}, \bibinfo {author} {\bibfnamefont {G.}~\bibnamefont
  {Christou}}, \bibinfo {author} {\bibfnamefont {V.~S.}\ \bibnamefont {Zapf}},
  \bibinfo {author} {\bibfnamefont {G.~K.-L.}\ \bibnamefont {Chan}},\ and\
  \bibinfo {author} {\bibfnamefont {H.-P.}\ \bibnamefont {Cheng}},\ }\bibfield
  {title} {\bibinfo {title} {Using hyperoptimized tensor networks and
  first-principles electronic structure to simulate the experimental properties
  of the giant {Mn}$_{84}$ torus},\ }\href
  {https://doi.org/10.1021/acs.jpclett.2c00354} {\bibfield  {journal} {\bibinfo
   {journal} {The Journal of Physical Chemistry Letters}\ }\textbf {\bibinfo
  {volume} {13}},\ \bibinfo {pages} {2365} (\bibinfo {year}
  {2022})}\BibitemShut {NoStop}%
\bibitem [{\citenamefont {Hagym\'asi}\ \emph {et~al.}(2021)\citenamefont
  {Hagym\'asi}, \citenamefont {Sch\"afer}, \citenamefont {Moessner},\ and\
  \citenamefont {Luitz}}]{Hagymasi2021}%
  \BibitemOpen
  \bibfield  {author} {\bibinfo {author} {\bibfnamefont {I.}~\bibnamefont
  {Hagym\'asi}}, \bibinfo {author} {\bibfnamefont {R.}~\bibnamefont
  {Sch\"afer}}, \bibinfo {author} {\bibfnamefont {R.}~\bibnamefont
  {Moessner}},\ and\ \bibinfo {author} {\bibfnamefont {D.~J.}\ \bibnamefont
  {Luitz}},\ }\bibfield  {title} {\bibinfo {title} {Possible inversion symmetry
  breaking in the $s=1/2$ pyrochlore heisenberg magnet},\ }\href
  {https://doi.org/10.1103/PhysRevLett.126.117204} {\bibfield  {journal}
  {\bibinfo  {journal} {Phys. Rev. Lett.}\ }\textbf {\bibinfo {volume} {126}},\
  \bibinfo {pages} {117204} (\bibinfo {year} {2021})}\BibitemShut {NoStop}%
\bibitem [{\citenamefont {Hagym\'asi}\ \emph {et~al.}(2022)\citenamefont
  {Hagym\'asi}, \citenamefont {Noculak},\ and\ \citenamefont
  {Reuther}}]{Hagymasi2022}%
  \BibitemOpen
  \bibfield  {author} {\bibinfo {author} {\bibfnamefont {I.}~\bibnamefont
  {Hagym\'asi}}, \bibinfo {author} {\bibfnamefont {V.}~\bibnamefont
  {Noculak}},\ and\ \bibinfo {author} {\bibfnamefont {J.}~\bibnamefont
  {Reuther}},\ }\bibfield  {title} {\bibinfo {title} {Enhanced
  symmetry-breaking tendencies in the $s=1$ pyrochlore antiferromagnet},\
  }\href {https://doi.org/10.1103/PhysRevB.106.235137} {\bibfield  {journal}
  {\bibinfo  {journal} {Phys. Rev. B}\ }\textbf {\bibinfo {volume} {106}},\
  \bibinfo {pages} {235137} (\bibinfo {year} {2022})}\BibitemShut {NoStop}%
\bibitem [{\citenamefont {Rausch}\ \emph {et~al.}(2021)\citenamefont {Rausch},
  \citenamefont {Plorin},\ and\ \citenamefont {Peschke}}]{Rausch2021}%
  \BibitemOpen
  \bibfield  {author} {\bibinfo {author} {\bibfnamefont {R.}~\bibnamefont
  {Rausch}}, \bibinfo {author} {\bibfnamefont {C.}~\bibnamefont {Plorin}},\
  and\ \bibinfo {author} {\bibfnamefont {M.}~\bibnamefont {Peschke}},\
  }\bibfield  {title} {\bibinfo {title} {{The antiferromagnetic $S=1/2$
  Heisenberg model on the C$_{60}$ fullerene geometry}},\ }\href
  {https://doi.org/10.21468/SciPostPhys.10.4.087} {\bibfield  {journal}
  {\bibinfo  {journal} {SciPost Phys.}\ }\textbf {\bibinfo {volume} {10}},\
  \bibinfo {pages} {087} (\bibinfo {year} {2021})}\BibitemShut {NoStop}%
\bibitem [{\citenamefont {Rausch}\ \emph {et~al.}(2022)\citenamefont {Rausch},
  \citenamefont {Peschke}, \citenamefont {Plorin},\ and\ \citenamefont
  {Karrasch}}]{Rausch2022}%
  \BibitemOpen
  \bibfield  {author} {\bibinfo {author} {\bibfnamefont {R.}~\bibnamefont
  {Rausch}}, \bibinfo {author} {\bibfnamefont {M.}~\bibnamefont {Peschke}},
  \bibinfo {author} {\bibfnamefont {C.}~\bibnamefont {Plorin}},\ and\ \bibinfo
  {author} {\bibfnamefont {C.}~\bibnamefont {Karrasch}},\ }\bibfield  {title}
  {\bibinfo {title} {{Magnetic properties of a capped kagome molecule with 60
  quantum spins}},\ }\href {https://doi.org/10.21468/SciPostPhys.12.5.143}
  {\bibfield  {journal} {\bibinfo  {journal} {SciPost Phys.}\ }\textbf
  {\bibinfo {volume} {12}},\ \bibinfo {pages} {143} (\bibinfo {year}
  {2022})}\BibitemShut {NoStop}%
\bibitem [{\citenamefont {Szab\'o}\ \emph {et~al.}(2024)\citenamefont
  {Szab\'o}, \citenamefont {Capponi},\ and\ \citenamefont {Alet}}]{Szabo2024}%
  \BibitemOpen
  \bibfield  {author} {\bibinfo {author} {\bibfnamefont {A.}~\bibnamefont
  {Szab\'o}}, \bibinfo {author} {\bibfnamefont {S.}~\bibnamefont {Capponi}},\
  and\ \bibinfo {author} {\bibfnamefont {F.}~\bibnamefont {Alet}},\ }\bibfield
  {title} {\bibinfo {title} {Noncoplanar and chiral spin states on the way
  towards n\'eel ordering in fullerene heisenberg models},\ }\href
  {https://doi.org/10.1103/PhysRevB.109.054410} {\bibfield  {journal} {\bibinfo
   {journal} {Phys. Rev. B}\ }\textbf {\bibinfo {volume} {109}},\ \bibinfo
  {pages} {054410} (\bibinfo {year} {2024})}\BibitemShut {NoStop}%
\bibitem [{\citenamefont {Rausch}\ \emph {et~al.}(2023)\citenamefont {Rausch},
  \citenamefont {Peschke}, \citenamefont {Plorin}, \citenamefont {Schnack},\
  and\ \citenamefont {Karrasch}}]{Rausch2023}%
  \BibitemOpen
  \bibfield  {author} {\bibinfo {author} {\bibfnamefont {R.}~\bibnamefont
  {Rausch}}, \bibinfo {author} {\bibfnamefont {M.}~\bibnamefont {Peschke}},
  \bibinfo {author} {\bibfnamefont {C.}~\bibnamefont {Plorin}}, \bibinfo
  {author} {\bibfnamefont {J.}~\bibnamefont {Schnack}},\ and\ \bibinfo {author}
  {\bibfnamefont {C.}~\bibnamefont {Karrasch}},\ }\bibfield  {title} {\bibinfo
  {title} {{Quantum spin spiral ground state of the ferrimagnetic sawtooth
  chain}},\ }\href {https://doi.org/10.21468/SciPostPhys.14.3.052} {\bibfield
  {journal} {\bibinfo  {journal} {SciPost Phys.}\ }\textbf {\bibinfo {volume}
  {14}},\ \bibinfo {pages} {052} (\bibinfo {year} {2023})}\BibitemShut
  {NoStop}%
\bibitem [{\citenamefont {Hubig}\ \emph {et~al.}(2015)\citenamefont {Hubig},
  \citenamefont {McCulloch}, \citenamefont {Schollw\"ock},\ and\ \citenamefont
  {Wolf}}]{Hubig2015}%
  \BibitemOpen
  \bibfield  {author} {\bibinfo {author} {\bibfnamefont {C.}~\bibnamefont
  {Hubig}}, \bibinfo {author} {\bibfnamefont {I.~P.}\ \bibnamefont
  {McCulloch}}, \bibinfo {author} {\bibfnamefont {U.}~\bibnamefont
  {Schollw\"ock}},\ and\ \bibinfo {author} {\bibfnamefont {F.~A.}\ \bibnamefont
  {Wolf}},\ }\bibfield  {title} {\bibinfo {title} {Strictly single-site dmrg
  algorithm with subspace expansion},\ }\href
  {https://doi.org/10.1103/PhysRevB.91.155115} {\bibfield  {journal} {\bibinfo
  {journal} {Phys. Rev. B}\ }\textbf {\bibinfo {volume} {91}},\ \bibinfo
  {pages} {155115} (\bibinfo {year} {2015})}\BibitemShut {NoStop}%
\bibitem [{\citenamefont {Ummethum}\ \emph {et~al.}(2013)\citenamefont
  {Ummethum}, \citenamefont {Schnack},\ and\ \citenamefont
  {Läuchli}}]{Ummethum2013}%
  \BibitemOpen
  \bibfield  {author} {\bibinfo {author} {\bibfnamefont {J.}~\bibnamefont
  {Ummethum}}, \bibinfo {author} {\bibfnamefont {J.}~\bibnamefont {Schnack}},\
  and\ \bibinfo {author} {\bibfnamefont {A.~M.}\ \bibnamefont {Läuchli}},\
  }\bibfield  {title} {\bibinfo {title} {Large-scale numerical investigations
  of the antiferromagnetic heisenberg icosidodecahedron},\ }\href
  {https://doi.org/https://doi.org/10.1016/j.jmmm.2012.09.037} {\bibfield
  {journal} {\bibinfo  {journal} {Journal of Magnetism and Magnetic Materials}\
  }\textbf {\bibinfo {volume} {327}},\ \bibinfo {pages} {103} (\bibinfo {year}
  {2013})}\BibitemShut {NoStop}%
\bibitem [{\citenamefont {Zauner-Stauber}\ \emph {et~al.}(2018)\citenamefont
  {Zauner-Stauber}, \citenamefont {Vanderstraeten}, \citenamefont {Fishman},
  \citenamefont {Verstraete},\ and\ \citenamefont
  {Haegeman}}]{Zauner-Stauber2018}%
  \BibitemOpen
  \bibfield  {author} {\bibinfo {author} {\bibfnamefont {V.}~\bibnamefont
  {Zauner-Stauber}}, \bibinfo {author} {\bibfnamefont {L.}~\bibnamefont
  {Vanderstraeten}}, \bibinfo {author} {\bibfnamefont {M.~T.}\ \bibnamefont
  {Fishman}}, \bibinfo {author} {\bibfnamefont {F.}~\bibnamefont
  {Verstraete}},\ and\ \bibinfo {author} {\bibfnamefont {J.}~\bibnamefont
  {Haegeman}},\ }\bibfield  {title} {\bibinfo {title} {Variational optimization
  algorithms for uniform matrix product states},\ }\href
  {https://doi.org/10.1103/PhysRevB.97.045145} {\bibfield  {journal} {\bibinfo
  {journal} {Phys. Rev. B}\ }\textbf {\bibinfo {volume} {97}},\ \bibinfo
  {pages} {045145} (\bibinfo {year} {2018})}\BibitemShut {NoStop}%
\bibitem [{\citenamefont {Lieb}\ and\ \citenamefont
  {Mattis}(1962)}]{Lieb_Mattis_1962}%
  \BibitemOpen
  \bibfield  {author} {\bibinfo {author} {\bibfnamefont {E.}~\bibnamefont
  {Lieb}}\ and\ \bibinfo {author} {\bibfnamefont {D.}~\bibnamefont {Mattis}},\
  }\bibfield  {title} {\bibinfo {title} {Ordering energy levels of interacting
  spin systems},\ }\href@noop {} {\bibfield  {journal} {\bibinfo  {journal}
  {Journal of Mathematical Physics}\ }\textbf {\bibinfo {volume} {3}},\
  \bibinfo {pages} {749} (\bibinfo {year} {1962})}\BibitemShut {NoStop}%
\bibitem [{\citenamefont {Schnack}\ \emph {et~al.}(2001)\citenamefont
  {Schnack}, \citenamefont {Schmidt}, \citenamefont {Richter},\ and\
  \citenamefont {Schulenburg}}]{Schnack2001}%
  \BibitemOpen
  \bibfield  {author} {\bibinfo {author} {\bibfnamefont {J.}~\bibnamefont
  {Schnack}}, \bibinfo {author} {\bibfnamefont {H.-J.}\ \bibnamefont
  {Schmidt}}, \bibinfo {author} {\bibfnamefont {J.}~\bibnamefont {Richter}},\
  and\ \bibinfo {author} {\bibfnamefont {J.}~\bibnamefont {Schulenburg}},\
  }\bibfield  {title} {\bibinfo {title} {Independent magnon states on magnetic
  polytopes},\ }\href {https://doi.org/10.1007/s10051-001-8701-6} {\bibfield
  {journal} {\bibinfo  {journal} {The European Physical Journal B - Condensed
  Matter and Complex Systems}\ }\textbf {\bibinfo {volume} {24}},\ \bibinfo
  {pages} {475} (\bibinfo {year} {2001})}\BibitemShut {NoStop}%
\bibitem [{\citenamefont {Schulenburg}\ \emph {et~al.}(2002)\citenamefont
  {Schulenburg}, \citenamefont {Honecker}, \citenamefont {Schnack},
  \citenamefont {Richter},\ and\ \citenamefont {Schmidt}}]{Schulenburg2002}%
  \BibitemOpen
  \bibfield  {author} {\bibinfo {author} {\bibfnamefont {J.}~\bibnamefont
  {Schulenburg}}, \bibinfo {author} {\bibfnamefont {A.}~\bibnamefont
  {Honecker}}, \bibinfo {author} {\bibfnamefont {J.}~\bibnamefont {Schnack}},
  \bibinfo {author} {\bibfnamefont {J.}~\bibnamefont {Richter}},\ and\ \bibinfo
  {author} {\bibfnamefont {H.-J.}\ \bibnamefont {Schmidt}},\ }\bibfield
  {title} {\bibinfo {title} {Macroscopic magnetization jumps due to independent
  magnons in frustrated quantum spin lattices},\ }\href
  {https://doi.org/10.1103/PhysRevLett.88.167207} {\bibfield  {journal}
  {\bibinfo  {journal} {Phys. Rev. Lett.}\ }\textbf {\bibinfo {volume} {88}},\
  \bibinfo {pages} {167207} (\bibinfo {year} {2002})}\BibitemShut {NoStop}%
\bibitem [{\citenamefont {Richter}\ \emph {et~al.}(2004)\citenamefont
  {Richter}, \citenamefont {Schulenburg}, \citenamefont {Honecker},
  \citenamefont {Schnack},\ and\ \citenamefont {Schmidt}}]{Richter2004}%
  \BibitemOpen
  \bibfield  {author} {\bibinfo {author} {\bibfnamefont {J.}~\bibnamefont
  {Richter}}, \bibinfo {author} {\bibfnamefont {J.}~\bibnamefont
  {Schulenburg}}, \bibinfo {author} {\bibfnamefont {A.}~\bibnamefont
  {Honecker}}, \bibinfo {author} {\bibfnamefont {J.}~\bibnamefont {Schnack}},\
  and\ \bibinfo {author} {\bibfnamefont {H.-J.}\ \bibnamefont {Schmidt}},\
  }\bibfield  {title} {\bibinfo {title} {Exact eigenstates and macroscopic
  magnetization jumps in strongly frustrated spin lattices},\ }\href
  {https://doi.org/10.1088/0953-8984/16/11/029} {\bibfield  {journal} {\bibinfo
   {journal} {Journal of Physics: Condensed Matter}\ }\textbf {\bibinfo
  {volume} {16}},\ \bibinfo {pages} {S779} (\bibinfo {year}
  {2004})}\BibitemShut {NoStop}%
\bibitem [{\citenamefont {Richter}(2005)}]{Richter2005}%
  \BibitemOpen
  \bibfield  {author} {\bibinfo {author} {\bibfnamefont {J.}~\bibnamefont
  {Richter}},\ }\bibfield  {title} {\bibinfo {title} {{Localized-magnon states
  in strongly frustrated quantum spin lattices}},\ }\href
  {https://doi.org/10.1063/1.2008130} {\bibfield  {journal} {\bibinfo
  {journal} {Low Temperature Physics}\ }\textbf {\bibinfo {volume} {31}},\
  \bibinfo {pages} {695} (\bibinfo {year} {2005})},\ \Eprint
  {https://arxiv.org/abs/https://pubs.aip.org/aip/ltp/article-pdf/31/8/695/8233353/695\_1\_online.pdf}
  {https://pubs.aip.org/aip/ltp/article-pdf/31/8/695/8233353/695\_1\_online.pdf}
  \BibitemShut {NoStop}%
\bibitem [{\citenamefont {Heinze}\ \emph {et~al.}(2024)\citenamefont {Heinze},
  \citenamefont {Kotte}, \citenamefont {Rausch}, \citenamefont {Demuer},
  \citenamefont {Luther}, \citenamefont {Feyerherm}, \citenamefont {Ammerlaan},
  \citenamefont {Zeitler}, \citenamefont {Gorbunov}, \citenamefont {Uhlarz},
  \citenamefont {Rule}, \citenamefont {Wolter}, \citenamefont {Kühne},
  \citenamefont {Wosnitza}, \citenamefont {Karrasch},\ and\ \citenamefont
  {Süllow}}]{heinze2024atacamitecu2cloh3highmagnetic}%
  \BibitemOpen
  \bibfield  {author} {\bibinfo {author} {\bibfnamefont {L.}~\bibnamefont
  {Heinze}}, \bibinfo {author} {\bibfnamefont {T.}~\bibnamefont {Kotte}},
  \bibinfo {author} {\bibfnamefont {R.}~\bibnamefont {Rausch}}, \bibinfo
  {author} {\bibfnamefont {A.}~\bibnamefont {Demuer}}, \bibinfo {author}
  {\bibfnamefont {S.}~\bibnamefont {Luther}}, \bibinfo {author} {\bibfnamefont
  {R.}~\bibnamefont {Feyerherm}}, \bibinfo {author} {\bibfnamefont {A.~A.
  L.~N.}\ \bibnamefont {Ammerlaan}}, \bibinfo {author} {\bibfnamefont
  {U.}~\bibnamefont {Zeitler}}, \bibinfo {author} {\bibfnamefont {D.~I.}\
  \bibnamefont {Gorbunov}}, \bibinfo {author} {\bibfnamefont {M.}~\bibnamefont
  {Uhlarz}}, \bibinfo {author} {\bibfnamefont {K.~C.}\ \bibnamefont {Rule}},
  \bibinfo {author} {\bibfnamefont {A.~U.~B.}\ \bibnamefont {Wolter}}, \bibinfo
  {author} {\bibfnamefont {H.}~\bibnamefont {Kühne}}, \bibinfo {author}
  {\bibfnamefont {J.}~\bibnamefont {Wosnitza}}, \bibinfo {author}
  {\bibfnamefont {C.}~\bibnamefont {Karrasch}},\ and\ \bibinfo {author}
  {\bibfnamefont {S.}~\bibnamefont {Süllow}},\ }\href
  {https://arxiv.org/abs/2410.01947} {\bibinfo {title} {Atacamite
  cu$_2$cl(oh)$_3$ in high magnetic fields: Quantum criticality and dimensional
  reduction of a sawtooth-chain compound}} (\bibinfo {year} {2024}),\ \Eprint
  {https://arxiv.org/abs/2410.01947} {arXiv:2410.01947 [cond-mat.str-el]}
  \BibitemShut {NoStop}%
\end{thebibliography}%
\end{document}